\documentclass[twocolumn,prl,showpacs,floatfix,altaffilletter]{revtex4-1}
\usepackage{amsmath,amssymb,amsfonts}
\usepackage{bm}
\usepackage{graphicx}
\usepackage{epsfig}
\usepackage{times}
\usepackage{color}
\usepackage[colorlinks,bookmarks=false,citecolor=blue,linkcolor=red,urlcolor=blue]{hyperref}

\newcommand{\fref}[1]{Fig.~\ref{#1}}
\newcommand{\eref}[1]{Eq.~(\ref{#1})}

\newcommand{\smallwidth}{0.6\columnwidth}
\newcommand{\figwidth}{0.97\columnwidth}

\begin{document}

\title{Model of the electron-phonon interaction and optical conductivity of Ba$_{1-x}$K$_x$BiO$_3$ superconductors}

\author{R.~Nourafkan$^{1}$, F.~Marsiglio$^{2}$, and G.~Kotliar$^{1}$}

\affiliation{$^{1}$ Department of Physics \& Astronomy, Rutgers University, Piscataway, NJ 08854-8019, USA}
\affiliation{$^{2}$ Department of Physics, University of Alberta, Edmonton, Alberta, Canada T6G 2E1}


\begin{abstract}
We investigate the physical properties of  the  Ba$_{1-x}$K$_x$BiO$_3$
compounds with a focus on the optical properties.
Results from the simple  Holstein model,  describing a single band
coupled to an oxygen breathing mode with parameters derived from
first principles calculations, are in excellent agreement
with a broad range of experimental information.
It accounts for  an insulating  parent compound at  $x=0$ 
with a direct- (optical) and an indirect-gap, and a metal
insulator transition around  $x = 0.38$. Strong electron-phonon
coupling leads to spectral weight redistribution  over a
frequency scale much larger than the characteristic phonon
frequency and to strongly anharmonic phonons. We find that the metallic
phase in the vicinity of phase boundary is close to the polaronic regime.
\end{abstract}

\maketitle
Compounds of the Ba$_{1-x}$K$_x$BiO$_3$ family exhibit both superconductivity
($T_c \approx 30$ K)
and an exotic semiconducting state \cite{Cava88,Sleight75, Varma88, Rice81, Bischofs02}.
The parent compound (BaBiO$_3$), with one valence electron per Bi atom,
is an insulator. Upon substitution of Ba with K an insulator-metal transition occurs at $x\approx 0.37$.
Significant effort has been made to understand optical conductivity (OC) 
measurements, which, in the metallic phase, display in addition to a Drude peak
a substantial amount of spectral weight in the
mid-infrared (MIR) frequency region \cite{Blanton93, Karlow93}.
However, there is no consensus on the origin of the features of the
OC. Interpretations have been put forward
arguing in favor of very weak e-ph  coupling \cite{Puchkov94, sharifi91, marsiglio96}
as well as very strong e-ph coupling, which can even be
responsible for polaronic and bipolaronic behavior 
\cite{Puchkov96, Bischofs02, Merz05,zasadzinski90, Hellman93, Tralshawala95}.

In this letter we  reevaluate the physical consequences of the
e-ph interaction in the Ba$_{1-x}$K$_x$BiO$_3$ family, within a
simple model with parameters  extracted from recent electronic structure
calculations \cite{Franchini09,Yin11}.  To solve the model
we use Dynamical Mean Field Theory (DMFT) \cite{Georges96}, 
with an Exact Diagonalization solver. We  focus on the impact of
e-ph coupling  on the OC, and  use comparison of
theory and  experiments  to elucidate the strength of the
e-ph coupling in this class of materials.

Band theory calculations indicate that a single anti-bonding Bi-O
$sp(\sigma)$ band crosses the Fermi level \cite{Mattheiss83,
Meregalli98}. Furthermore,  only optical phonons couple
strongly to the electrons. Our central approximation
is to retain the coupling
to a single phonon mode, the breathing mode, and take its
coupling constant to be independent of momentum (this derivation
is described in the supplemental materials (SM)). This mode is most
important since its condensation gives rise to the charge ordered
insulating state at low filling \cite{Inumaru08}.  Therefore we use a minimal single band model
Hamiltonian with Holstein e-ph coupling (see SM). 

A quantity of central importance is the  effective e-ph coupling
strength  defined by the low frequency behavior of the electron
self-energy. In DMFT, the effective mass of the electron is given by
$m_{\rm b}/m^*=\left( 1-\partial{\rm Re}\Sigma(\omega)/\partial
\omega|_{\omega=0} \right)^{-1}$, which is related to the
effective coupling, $\lambda_{\rm eff}$, by $m^*/m_{\rm
b}=1+\lambda_{\rm eff}$. $m_{\rm b}$ is the electronic band mass
of the carriers. The value of $\lambda_{\rm eff}$ includes phonon
renormalization effects, and will generally be significantly
higher than the bare coupling, $\lambda_0$, defined as
$(2g^2/\hbar \omega_0)N(\epsilon_F)$, where $g$ is e-ph coupling constant, 
$\omega_0$ is phonon frequency  and $N(\epsilon_F)$ is the 
density of states at the Fermi level. A second way to estimate
the effective  or renormalized value of lambda is via the formula,
$\lambda_{\rm eff} = -g^2 N(\epsilon_F){\rm Re}D(0)=\lambda_0/(1+2{\rm Re}\Pi(0)/\hbar\omega_0)$, where $D(\omega)$ denotes phonon propagator and $\Pi(\omega)$ is phonon self-energy.  
In our case the above definitions lead to the same value for $\lambda_{\rm eff}$. 
It is the renormalized value of the e-ph coupling,
i.e. $\lambda_{\rm eff}$, that should be compared with experiments
\cite{marsiglio90,Bauer11}.

In DMFT \cite{Georges96}, due to the absence of vertex corrections,
the OC can be expressed as a functional of
the fully interacting single particle Green function,
${\bf G}(\varepsilon_{\bf k},\hbar \omega)$,
as  

\begin{eqnarray}
{\rm Re}\ \sigma_{\alpha \alpha}(\omega)&=&(1+\frac{F_1}{3})\frac{(\pi e^2/\hbar)}{V_{cell}N}\sum_{{\bf k} , \sigma}
\int d\epsilon \ \frac{f(\epsilon)-f(\hbar\omega+\epsilon)}{\hbar\omega} \nonumber \\
& \times&{\rm Tr}\big[{\bf v}_{\alpha}({\bf k})A_{{\bf k}\sigma}(\epsilon){\bf v}_{\alpha}({\bf k})A_{{\bf k}\sigma}(\hbar\omega+\epsilon)\big],
\label{eq:Optical}
\end{eqnarray}
where $A_{{\bf k}\sigma}(\epsilon)\equiv(-1/\pi){\rm Im}{\bf
G}_{\sigma}(\varepsilon_{\bf k},\epsilon)$ is the interacting
spectral function, ${\bf v}_{\alpha}({\bf k})\equiv
\frac{\partial {\bm \varepsilon}_{\bf k}}{\partial k_{\alpha}}$,
$f(\epsilon)$ is the Fermi function, and $V_{cell}=N_{\rm Bi}a^3$
is the unit cell volume. $N_{\rm Bi}$ is the number of Bi ions in
a unit cell. The lattice constant value is $a=4.27 \AA$. A Landau parameter $F_1$ \cite{Millis04}
renormalizes the current operator in \eref{eq:Optical} \cite{Landau}. 

When only intra-band optical transitions relative to the
lowest conduction band contribute to the optical spectral weight (OSW),
one obtains the \textit{restricted} or \textit{partial sum rule} \cite{maldague77}

\begin{eqnarray}
W_{\rm opt}&=&2\int_0^{\infty} d\omega \ {\rm Re}\sigma_{\alpha \alpha}(\omega)  \nonumber \\ &=& (1+\frac{F_1}{3})
\frac{\pi e^2}{\hbar^2 V_{cell}}\frac{1}{N}\sum_{{\bf k}\sigma} n_{{\bf k}\sigma}\frac{\partial^2 \epsilon ({\bf k})}{\partial k^2_{\alpha}}.
\label{eq:sumrule}
\end{eqnarray}
From a knowledge of the optical spectral weight, we can calculate the plasma frequency, given by
$W_{\rm opt}={\omega^2_p}/{4}$.

To appreciate the range of frequencies over which the optical
spectral weight redistributes, it is customary to define the
effective carrier number per Bi ions participating in optical
transitions, $N_{\rm eff}(\omega)$, defined by the partial
integral of the OC for frequencies less than
$\omega$ as
\begin{eqnarray}
\big(\frac{m_0}{m_{\rm b}} \big)N_{\rm Bi}N_{\rm eff}(\omega)&=&\frac{2m_0V_{cell}}{\pi e^2 }\int_0^{\omega} d\nu{\rm Re}\sigma_{\alpha \alpha} (\nu),
\end{eqnarray}
where
$m_0$ is the free-electron mass.  In order to compare the
theoretical results with the experiments of Ref. [\onlinecite{Karlow93}] for $N_{\rm
eff}(\omega)$ we adopt their convention and set $m_0/m_{\rm b}=1$.

Integrating up to infinite frequency one recovers the OSW.  A reduction of $W_{opt}$ relative to its non-interacting value, as given, for example, 
in a LDA estimation \cite{HSE}, is usually taken as a sign of the presence of
electron-electron correlation \cite{Basov05}. Before focusing on the Ba$_{1-x}$K$_x$BiO$_3$
compounds, we now revisit this issue on general grounds for the
e-ph problem. 

\begin{figure}
\begin{center}
\center{\includegraphics[width=\figwidth]{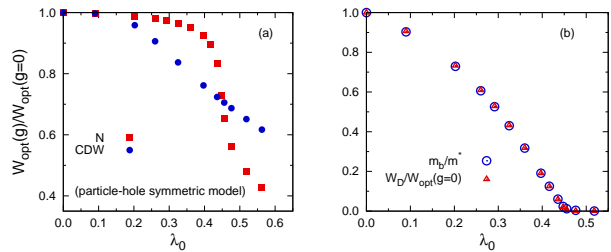}}
\caption{(Color online) Panel(a): normalized interacting OSW as a function of bare e-ph coupling, $\lambda_0$, at half-filling ($x=0$) for a particle-hole symmetric model in the normal and CDW phases. Panel(b): normalized Drude weight and inverse effective mass as a function of $\lambda_0$. The TB parameters for this figure are: $t_1 = 0.3926$ eV and $\hbar\omega_0=0.08$ eV.
}
\label{fig:sumrule}
\end{center}
\end{figure}

\fref{fig:sumrule} (a) displays
the normalized OSW in the normal and CDW phases at half-filling ($x=0$) for a
particle-hole symmetric model. 
The most important feature of \fref{fig:sumrule} (a)
is that in the weak to intermediate range of $\lambda_0$ the
total OSW in a homogeneous normal metallic phase
depends only weakly on the strength of the
e-ph coupling as in Migdal-Eliashberg theory. The OSW reduction due to e-ph coupling  is only appreciable by approaching the polaronic
regime, but even upon entering that regime the OSW is substantial while the effective mass is considerably
enhanced, as shown in \fref{fig:sumrule} (a) and (b). 
These results are similar away from half-filling.
\fref{fig:sumrule} (b) shows the Drude part \cite{Drude} normalized
by the non-interacting total optical integral, $W_{\rm opt}(g=0)$. In our case, without any
other interaction, $W_{\rm opt}(g=0)=W_D(g=0)$. 
The $W_D(g)/W_D(g=0)$ and inverse effective mass agree and both go continuously to zero at the critical coupling strength, $\lambda_{0c} \approx 0.455$.

Hence the variation in the Drude weight or the effective mass with e-ph coupling is more
prominent than the variation in OSW.
\fref{fig:sumrule} (a) also shows that for $\lambda_0 < \lambda_{0c}$, ordering of
itinerant carriers in a charge density wave (CDW) phase leads to a larger reduction of OSW than normal phase.
However, for $\lambda_0 \ge \lambda_{0c}$, the carriers are localized in real space, due to
bipolaron formation, and CDW ordering enhances OSW. Therefore, the e-ph coupling dependence
of the difference between OSW in the metallic and
CDW phases is not monotonic (similar to the temperature dependence of the
OSW \cite{Ciuchi08}).

To  focus on the Ba$_{1-x}$K$_x$BiO$_3$ compounds, we determine
the hopping parameters to fit  the LDA band structure
\cite{Mattheiss83} which results in following hopping parameters
$t_1=0.3926, t_2=0.0516, t_3=-0.0017, t_4=-0.0987$, all in eV. 
The strong energy variation of the electron DOS for the
real material (see SM) leads to a strong doping dependence of  the bare
e-ph coupling, $\lambda_0$, which  plays a central  role for
understanding the physical properties of the
Ba$_{1-x}$K$_x$BiO$_3$ family. For the phonon frequency and e-ph coupling we choose a set of the
{\em bare} parameters, $\hbar\omega_0=80$ meV, and $g=0.2275$ eV.
The value of the e-ph
coupling to the breathing mode, $g$,  matches closely to the recent results of
first principles calculation of the deformation potential
at the zone boundary \cite{Yin11} (see SM). 
The characteristic phonon frequency
represents a {\em bare} parameter. 
Below, we shall see that it becomes renormalized, 
so that the final value is representative of the experimental value.
For simplicity we take all the bare parameters in the Hamiltonian
to be independent of doping (rigid band picture). 
Considering the simplifications made, our results are in surprisingly good  agreement with
experiments.

\begin{figure}
\begin{center}
\includegraphics[width=\figwidth]{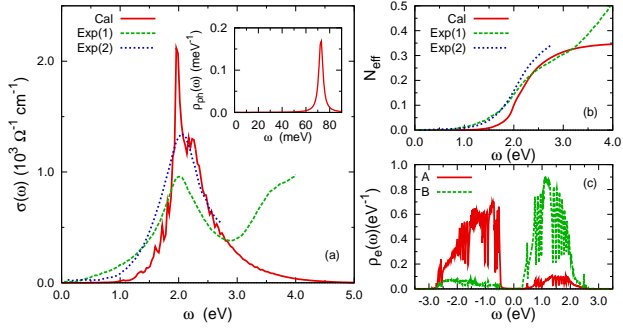}
\caption{(Color online) (a) Calculated and measured OC at half-filling ($x=0$).
Measured OCs of BaBiO$_3$ are reproduced from \cite{Karlow93,Blanton93}, labeled by $(1)$ and $(2)$, respectively.
Inset: The renormalized phonon density of states.
(b) Calculated and measured effective carrier number per Bi ion participating in the optical transitions.
(c) Electron DOS in the charge ordered phase with an indirect gap of the order of $\Delta\approx 0.5$ eV. A and B indicate
the majority and minority sublattices for electron density.}
\label{fig:OpticsCO}
\end{center}
\end{figure}

For the parent compound, BaBiO$_3$ ($x=0$), we show the calculated conductivity as a function of
frequency in panel (a) of \fref{fig:OpticsCO}, along with
two measurements \cite{Karlow93,Blanton93} performed at room temperature.
The calculated conductivity displays a characteristic insulating spectrum; the Drude peak is absent, and
spectral weight has been transferred to
the region above the gap. The calculated gap value, $ \Delta_{\rm opt}\approx 2$ eV, and the subsequent absorption
strength are similar in magnitude to the observed values  \cite{Karlow93,Blanton93}. 
The inset shows the
phonon density of states in CDW phase with a renormalized peak near $\approx 72$ meV, in good agreement with the 
experimental value of the breathing mode phonon frequency (Experimental data in \cite{Meregalli98}) \cite{pol}.

In \fref{fig:OpticsCO} (b) we show
$N_{\rm eff}(\omega)$ vs. $\omega$ for the undoped system. The calculated $N_{\rm eff}$ is in good agreement
with the experimental data in the range of frequencies in which inter-band
transitions play no role ($\omega \le 3$ eV). \fref{fig:OpticsCO} (c) shows the electron DOS in the charge ordered phase; it  shows a gap value
$\approx 0.5 {\rm eV}$, which corresponds to the indirect gap in BaBiO$_3$ \cite{Takagi87}.

Upon doping the system we find that at doping value around $x \approx 0.38$,
charges distribute
uniformly between the two sub-lattices. Investigation of the electronic DOS
indicates that this is a transition to a metallic state with a finite DOS at the Fermi energy. In the
intermediate doping range $0<x<0.38$, the electron density oscillates between two values in successive
iterations of the DMFT cycle and a converged answer was not found. This implies that the system either has a tendency towards another phase, like a non-commensurate CDW, or a tendency towards phase separation,
both of which cannot be seen in our study \cite{Freericks94}.

\begin{figure}
\begin{center}
\center{\includegraphics[width=\figwidth]{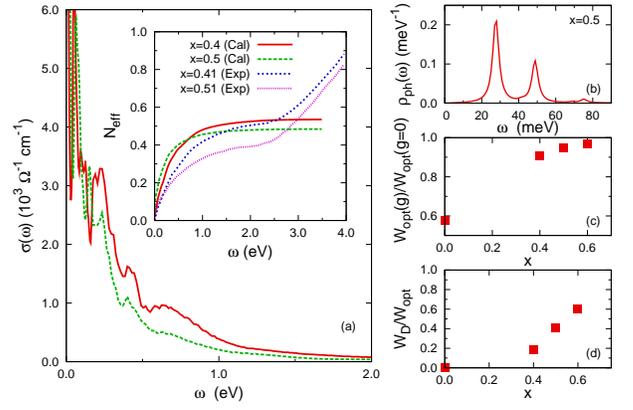}}
\caption{(Color online) (a) Calculated and experimental \cite{Karlow93}
conductivity for the indicated doping $x$
at zero temperature.
Inset:
Effective carrier number participating in the OC
as a function of frequency.  
(b) The phonon density of states at $x=0.5$.   
(c) Normalized interacting OSW by non-interacting value as a function of doping.
(d) Drude weight, $W_{\rm D}$,
of the OC, normalized by total weight $W_{\rm opt}$,
as a function of the doping level. In panels (c) and (d), system at $x=0$ is in CDW phase, while other data points show normal metallic phase.}
\label{fig:OpticsM}
\end{center}
\end{figure}

\fref{fig:OpticsM} (a) shows the calculated OC
for two doping values in the metallic phase.
As the system is doped into the metallic phase, the optical response shows a Drude peak,
centered at $\omega=0$.
This indicates coherent quasi-particle motion,
and an additional mid-infrared (MIR) feature arises, caused by single- and multi-phonon excitations
and absorption processes. The MIR contribution is an incoherent component of the OC,
and corresponds to optical transitions in which the carrier
is excited but leaves behind a local distortion, which corresponds to real excitations.
This MIR component occurs at frequencies characteristic of these excitations.
As we saw in \fref{fig:sumrule} (b), the e-ph interaction causes a reduction of the free carrier
Drude peak weight; in fact it is approximately reduced by the
quasi-particle renormalization factor  $(1 + \lambda_{\rm eff})^{-1}$ and the
spectral weight increases in the MIR energy range
reflecting the incoherent scattering excitations that result from absorption
processes assisted by the e-ph interaction.
The MIR frequency range extends well beyond that of the characteristic phonon frequency.

In the inset of  \fref{fig:OpticsM} we show the effective carrier number calculated from OC. In the metallic phase $N_{\rm eff}(\omega)$ at first increases steeply at low frequencies, on account
of the appearance of the Drude band centered at $\omega=0$, and continues to increase in the
mid-infrared region, eventually saturating at a high frequency.
The trends are in good agreement with the experimental data \cite{Karlow93, Neff}.

\fref{fig:OpticsM} (b) shows the phonon density of states. The bare phonon frequency is softened and 
the phonon DOS shows structures similar to those observed experimentally \cite{Huang90}. The effective oscillator potential and anharmonicity is further explored in the SM. Consistent with experimental results \cite{Menushenkov00}, the effective potential shows a double well structure, with one minimum disappearing with increased doping.

We now turn to the doping dependence of the OSW. Due
to a higher value of the DOS at  ($x=0$)  the  e-ph
coupling ($\lambda_0$) is largest in the parent compound and decreases with
doping.
\fref{fig:OpticsM} (c) shows the normalized interacting OSW 
vs. doping.  It is clear that the OSW is greatly reduced for small values of $x$ due to the e-ph interaction.
In the metallic phase, upon increasing
the doping level, electrons become more undressed, which leads
to a smaller optical weight reduction.

\fref{fig:OpticsM} (d) shows the Drude part (the MIR part is $W_{\rm opt} - W_D$)
normalized by the total optical integral
as a function of doping at $T=0$ \cite{sumrule}.
In the metallic phase the OSW redistributes
between the Drude peak and the MIR shoulder, and by
increasing the doping level the Drude weight rises very
quickly.
For instance, the Drude peak weight increased by almost a factor of two
with increasing $x$ from $x=0.4$ to $x=0.5$.

In the DMFT framework, the effective e-ph coupling can be extracted either from the phonon propagator at zero frequency, or the effective mass, or from the relative weight of the Drude peak to the total OSW in the absence of e-ph coupling, $ W_{\rm D}(g)/W_{\rm opt}(g=0)\approx (1+\lambda_{\rm eff})^{-1}$.
We found that  the effective e-ph coupling is strongly doping dependent
with the highest value $(\lambda_{\rm eff} \ge 4)$ at  $x=0.4 $ close to the transition to the insulating phase.
With such a high value of the effective e-ph coupling the formation of an inhomogeneous phase
in the vicinity of the phase boundary is highly likely.
At higher doping levels where the system is likely to
be homogeneous, we find $\lambda_{\rm eff} \approx 1.6$ for $x=0.5$.

The calculated plasma frequency for $x=0.5$ is $\hbar^2\omega_p^2=8.95 \ {\rm eV}^2$,
again in good agreement with the experimental value \cite{Hellman93}.
Using the calculated values for the plasma frequency for $x=0.5$
we can use the  high temperature behavior of the measured resistivity to
estimate the e-ph coupling, $\lambda_{\rm eff}^{(\rm tr)}$ \cite{remark1}.
Using the high temperature resistivity formula \cite{Tralshawala95},
\begin{equation}
\rho(T)\approx \rho_{\rm ph}(T)\approx \frac{8\pi^2 k_B}{\hbar}\frac{\lambda_{\rm eff}^{(\rm tr)}}{\omega_p^2}T
\label{eq:rho}
\end{equation}
we estimate a value of $\lambda_{\rm eff}^{(\rm tr)}\approx 1.75$.

In support of strong e-ph coupling, heat capacity measurements for Ba$_{1-x}$K$_x$BiO$_3$ with $x\approx 0.4$ show a jump in
the specific heat of the order $\Delta C_{p}(T_c)/T_c \approx 4-5
\ {\rm mJ} \ {\rm mol}^{-1}{\rm K}^{-2}$ at  $T_c$ \cite{Blanchard02}. One
can relate this to the coefficient of the electronic specific heat,
$\gamma$, by the BCS expression $\Delta C_{p}(T_c)/T_c=1.43\gamma$ \cite{remark2} 
. Then from the relation $\gamma=(2/3)\pi^2N(\epsilon_F)k_B^2(1+\lambda_{\rm
eff})$  with use of $N(\epsilon_F)=0.23 \ {\rm states}/{\rm eV} \ {\rm spin} \ {\rm cell}$ for $x=0.5$,
$\lambda_{\rm eff}$ is estimated to be $1.6-2.2$, consistent with our previous estimates.

In conclusion we have revisited the physical properties of
Ba$_{1-x}$K$_x$BiO$_3$,  using a model Hamiltonian with parameters
inspired by recent first principles calculations \cite{Franchini09, Yin11} 
but going beyond these calculations by  taking into
account the full many body physics and inherent anharmonicity of
a strongly coupled e-ph system. For small
doping the coupling is strong enough to sustain  polaron and small bipolaron
formation and the coupling decreases for large doping.  For the
metallic regime near the phase transition boundary, we find a situation
intermediate between a Fermi/BCS liquid and a bipolaronic  superconductor, in agreement with
experimental studies \cite{Uemura91, Homes06}.  
The model accounts for experimentally
determined gaps (both direct and indirect) in the parent compound
at $x=0$ and an abrupt insulator transition at a critical doping.
In the metallic phase, we find a strongly anharmonicity  in
agreement with recent experiments \cite{Menushenkov00} and transfer of substantial electron phonon coupling strength
to low frequencies as shown in the phonon density of
states. 
Strong e-ph coupling reduces the width of the Drude
peak and transfers spectral weight to the mid-infrared structure.
Spectral weight redistributes over a frequency scale that is much
larger than the
phonon frequency. 
Further extensions will require  the incorporation of couplings
to other phonon modes and of additional bands in the modeling.
These are not expected to change drastically  the  main
conclusions presented in this paper which focused on the
electronic properties  but are certainly required for a more realistic
study of the phonon spectra.

\paragraph{Acknowledgments.}
This research was supported by the AFOSR MURI
“Towards New and Better High Temperature Superconductors.” GK
is grateful to Dimitri Basov for useful  discussions.  RN and GK
would like to thank Zhiping Yin for many helpful discussion. FM
acknowledges support by the Natural Sciences and Engineering
Research Council of Canada (NSERC). FM and GK acknowledge the
support of the Canadian Institute for Advanced Research (CIfAR).

\bibliographystyle{prsty}

\begin{thebibliography}{10}
\bibitem{Landau}
The introduction of the $F_1$ can be
motivated as follows; Hybrid density functional theory contains
about twenty  to thirty percent exchange
contributions to the band dispersions, but exchange contributions
should not contribute to the current via the Peierls substitution.
We incorporate this effect via a Landau coefficient $(1+F_1/3)$
multiplying the Peierls expression relating  the current to the
derivative of the energy dispersion with respect to momentum. We
adopt a value of $ 1 + F_1/3 = 0.8$, removing twenty percent of
the current originating from the exchange term in the self
energy.
\bibitem{HSE}
The HSE values for optical spectral weights in the metal are close to the LDA value.
\bibitem{Drude}
The Drude weight,
$W_D$, is determined by examining the asymptotic form of the current-current susceptibility
on the imaginary axis \cite{Scalapino93}.
\bibitem{pol}
The e-ph
coupling in the parent compound is sufficiently strong to
cause real-space pairing of carriers in the form of small
bipolarons; however, this phase is hidden by the charge ordered
phase.
\bibitem{Neff}
Indeed, upon increasing the doping level one expects that $(m_0/m_{\rm b})N_{\rm eff}(\omega_c)$ gets closer to the
expected carrier number determined from the chemical composition $1-x$. Here, $\hbar\omega_c$ is a cutoff energy which must be
chosen to exhaust the oscillator strength of the absorption in question but lie below the rest of the absorption bands.
Assuming a small variation in $(m_0/m_{\rm b})$ between $x=0.4-0.5$, then the difference between $N_{\rm eff}(\omega_c)$ and $1-x$ decreases upon increasing the doping level. Our calculations show this behavior. While the early experimental data of Ref. [\onlinecite{Karlow93}] shows the opposite trend, later observation shows a trend consistent with our
calculated results \cite{Puchkov96}.
\bibitem{sumrule}
The total weight, $W_{\rm opt}$,
determined by summing the area under the conductivity,
agrees with the calculated partial sum rule,
which is an independent calculation (right hand side of \eref{eq:sumrule}),
and confirms the sum rule.
\bibitem{remark1} The superscript `tr' is for transport properties. For our purposes we will regard this as equivalent to the effective coupling strength defined through the effective mass.
\end{thebibliography}

\newpage
\twocolumngrid
\vspace{0.2in}
\begin{center}
{\bf \large Model of the electron-phonon interaction and optical conductivity of Ba$_{1-x}$K$_x$BiO$_3$ superconductors - Supplemental materials}
\end{center}
\vspace{0.1in}

\renewcommand{\thetable}{S\Roman{table}}
\renewcommand{\thefigure}{S\arabic{figure}}
\renewcommand{\thesubsection}{S\arabic{subsection}}
\renewcommand{\theequation}{S\arabic{equation}}

\setcounter{secnumdepth}{1}
\setcounter{equation}{0}
\setcounter{figure}{0}
\setcounter{section}{0}

\section{Electron-phonon coupling}
Here we provide additional information on the derivation of a Holstein-type e-ph coupling from 
the Rice-Sneddon model Hamiltonian \cite{Rice1981}. The phonon part and e-ph interaction in the Rice-Sneddon 
model Hamiltonian is given by

\begin{multline}
H_{ph}+H_{e-ph}=\frac{M_{\rm O}}{2}\sum_{i\alpha}(\dot{u}^2_{i\alpha}+
\omega_0^2u^2_{i\alpha})\\-\mathcal{G} \sum_{i\alpha} n_i(u_{i\alpha +}-u_{i\alpha -})
\tag{S1}\label{eq:RS}
\end{multline} 

where $u_{i\alpha \pm}$ denote displacements of two oxygen ions 
at sites ${\bf R}_i \pm a \hat {{\bf e}}_{\alpha}/2$, 
placed in the $\alpha$th direction at a distance $a/2$, 
and $\mathcal{G}$ is the strength of this interaction. 
$M_{\rm O}$ is the oxygen ion mass, and $\omega_0$ is the phonon frequency. 

Introducing the  boson representation 
for each oxygen coordinate, 
$u_{i\alpha}=\sqrt{\hbar/2M_{\rm O}\omega_0}(B^{\dagger}_{i\alpha}+B_{i\alpha})$, Eq. (\ref{eq:RS}) reads

\begin{multline}
H_{ph}+H_{e-ph}=\hbar\omega_0\sum_{i\alpha}(B^{\dagger}_{i\alpha}B_{i\alpha}
+\frac{1}{2}) \\-\mathcal{G}\sqrt{\frac{\hbar}{2M_{\rm O}\omega_0}}\sum_{i\alpha}n_{i}
[(B^{\dagger}_{i\alpha+}-B^{\dagger}_{i\alpha-})+h.c.].
\tag{S2}\label{eq:Ham1}
\end{multline}

With these phonon coordinates, being attached to oxygen ions, 
there is a nonzero overlap between nearest-neighbor operators, i.e., 
they are not orthogonal. It is better to define new orthogonal boson operators \cite{Piekarz00}. 
By Fourier transforming the boson operators, 
$B_{i\alpha \pm}=(1/\sqrt{N})\sum_{\bf k}\exp[i{\bf k}.({\bf R}_i \pm a \hat {{\bf e}}_{\alpha}/2)]B_{{\bf k}\alpha \pm}$, it is easy to show

\begin{equation}
H_{e-ph}=-\frac{2\mathcal{G}}{\sqrt{N}}\sqrt{\frac{\hbar}{2M_{\rm O}\omega_0}}\sum_{i}n_{i}
\sum_{{\bf k} }\mu_{\bf k}(e^{i{\bf k}.{\bf R}_i}b^{\dagger}_{{\bf k}}+h.c.),
\tag{S3}\label{eq:Ham2}
\end{equation} 
where $N$ is the number of sites, $b_{{\bf k}}$ represents a symmetric phononic mode 

\begin{equation}
b_{{\bf k}} = -\frac{i}{\mu_{\bf k}}\sum_{\alpha} B_{{\bf k}\alpha} sin\frac{k_{\alpha}}{2},
\tag{S4}\label{eq:boson_Op}
\end{equation}
and 

\begin{equation}
\mu_{\bf k}=\sqrt{sin^2(k_x/2)+sin^2(k_y/2)+sin^2(k_z/2)}.
\tag{S5}\label{eq:Mu}
\end{equation}
The new operators, $b_{{\bf k}}$, fulfill the boson commutation relations. 
By transforming the new operator back to real space 
we obtain the following form of the phonon related terms

\begin{multline}
H_{ph}+H_{e-ph} = \hbar \omega_0 \sum_i (b^{\dagger}_{i}b_i+1/2) \\-\sum_{ij} g_{ij}n_i(b^{\dagger}_{j}+b_j) 
\tag{S6}\label{eq:Ham3}
\end{multline} 

where $g_{ij}=2\mathcal{G} \sqrt{\hbar/2M_{\rm O}\omega_0}\gamma_{ij}$. 
The coefficients 
$\gamma_{ij}=\frac{1}{\sqrt{N}}\sum_{\bf k} \mu_{\bf k} \exp[-i{\bf k}.({\bf R}_i-{\bf R}_j)]$ 
are responsible for the long-range e-ph interaction, which is a 
consequence of orthogonalization of the local phonon modes. 
However, these terms fall off rapidly with distance 
(e.g. $\gamma(0)=1.19$, and $\gamma(1)=-0.11$) \cite{Piekarz00}. In the Holstein model, we only keep the first term 
$g\equiv g_{ii}=2\mathcal{G} \sqrt{\hbar/2M_{\rm O}\omega_0}\gamma(0)$. Therefore, we use following model Hamiltonian:

\begin{multline}
H=-\sum_{ i,j, \sigma} t_{ij}(c_{i\sigma}^{\dagger}c_{j\sigma} + h.c.)
\\ + g\sum_i (n_i-1) (b_i + b_i^{\dagger}) + \hbar\omega_0 \sum_i b_i^{\dagger}b_i ,
\tag{S7}\label{eq:Holstein}
\end{multline}

where $c_{i\sigma}$ ($c_{i\sigma}^{\dagger}$) and  $b_{i}$
($b_{i}^{\dagger}$) are the destruction (creation) operators for
electrons with spin $\sigma$ and local vibrons with frequency
$\omega_0$, respectively. The operator $n_i \equiv \sum_\sigma
c_{i\sigma}^{\dagger} c_{i\sigma}$ is the electron density at
site $i$, the parameters $t_{ij}$ are the electron hopping matrix
elements between sites at $i$ and $j$, and $g$ denotes the e-ph coupling
constant. 

Using $\hbar \omega_0^{\rm LDA}=0.065$ eV,
and $\mathcal{G}=\mathcal{D}^{\rm LDA}/6$ the value of $g$ would be $g=0.237$ eV. Here $\mathcal{D}^{\rm LDA}=13.3\ {\rm eV}\AA^{-1}$ is the deformation potential obtained from the LDA calculation \cite{Yin2011}. 

The displacement of the oxygen ion is approximately given by $u_{i\alpha}\approx(1/3)\sqrt{\hbar/2M_{\rm O}\omega_0}\gamma(0)\langle b_i+b_i^{\dagger}\rangle$. Our model Hamiltonian calculation, with $\hbar \omega_0=0.08$ eV, gives a value $\langle b_i+b_i^{\dagger}\rangle\approx 4.5$ for the undoped case, which leads to a breathing mode distortion of the order of $u_i\approx 0.073 \AA$, in agreement with the experimental value  $0.085 \AA$  \cite{Cox76}.

\section{Density of states}

\begin{figure}
\begin{center}
\includegraphics[width=\smallwidth]{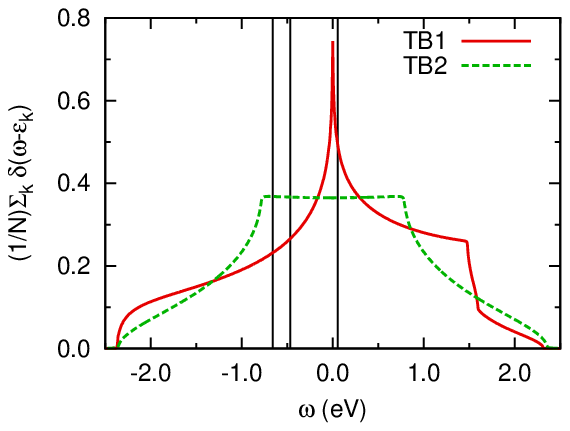}
\caption{(Color online) The density of states of the tight-binding model (TB1). Note the van Hove singularity. Vertical lines indicate the
non-interacting chemical potential for $x=0.5$, $x=0.4$ and $x=0$. Also shown is the DOS for nearest neighbor hopping only (TB2).}
\label{fig:dos}
\end{center}
\end{figure}

In \fref{fig:dos} we show the resulting density of state (DOS)
for a specific choice of hopping parameters $t_1=0.3926,
t_2=0.0516, t_3=-0.0017, t_4=-0.0987$, all in eV (TB1). These
correspond to successively further neighbor hopping. Note the
appearance of a (diverging) van Hove singularity, the result of
including beyond nearest neighbor hopping, and in strong contrast
to the rather flat DOS for the nearest neighbor model (TB2). The strong energy variation of the electron DOS for the
real material leads to a strong doping dependence of  the bare
e-ph coupling, $\lambda_0$, which  plays a central  role for
understanding the physical properties of the
Ba$_{1-x}$K$_x$BiO$_3$ family.

\section{Effective phonon potential}
In an electron-phonon system, an effective phonon potential, $V_{{\rm eff}}$, can be defined 
using an effective wavefunction defined by $\psi_{\rm eff} (q) = \sqrt{P(q)}$, 
which is taken to be a solution of the one dimensional Schrodinger equation \cite{Hewson10},

\begin{figure}
\begin{center}
\center{\includegraphics[width=\figwidth]{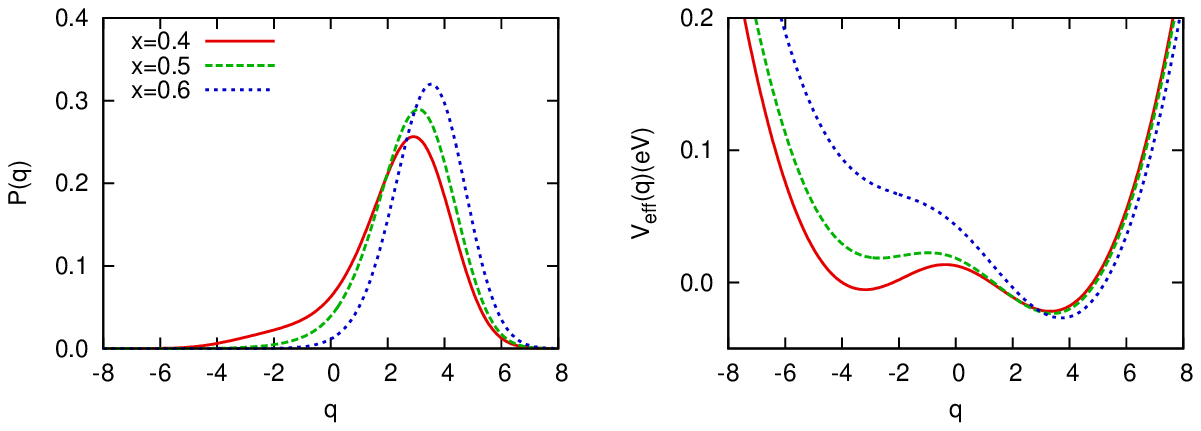}}
\caption{(Color online) The phonon displacement probability-distribution function (left panel) and the effective phonon potential (right panel) for three different doping levels. }
\label{fig:PhPDF}
\end{center}
\end{figure}

\begin{equation}
-\hbar\omega_0\frac{d^2\psi_{\rm eff} (q)}{dq^2}+V_{\rm eff}(q)\psi_{\rm eff} (q)=E\psi_{\rm eff} (q),
\tag{S8}
\end{equation}
where $\hat{q}\equiv b+b^{\dagger}$ and $P(q)=\langle \phi_0|q\rangle \langle q|\phi_0 \rangle$ 
is the phonon displacement probability-distribution function (PDF) \cite{Capone06}.  The wave function $|\phi_0\rangle$ is 
the ground-state wave function and $|q\rangle \langle q|$ is the projection operator 
onto the subspace where the phonon displacement operator, $\hat{q}$, at a given site has value $q$. 
This quantity is a measure of the distribution of the local distortions. 
Having $P(q)$, the effective phonon potential $V_{{\rm eff}}(q)$ is given by 

\begin{equation}
V_{\rm eff}(q) = E + \frac{\hbar\omega_0}{2}\left(\frac{P^{\prime \prime} (q)}{P (q)}-\frac{1}{2}\left( \frac{P^{\prime} (q)}{P (q)}\right)^2 \right).
\tag{S9}\label{eq:Veff}
\end{equation}

In \fref{fig:PhPDF} we show results for the probability distribution function and corresponding effective potential as deduced from \eref{eq:Veff} for three different doping levels (we ignore the constant $E$, so we only discuss the variation in the shape of the potential and not its absolute value). In each case $P(q)$ has a narrow peak which shifts to slightly larger $q$ with increasing doping. The average value of the displacement, $\langle q\rangle=\int_{-\infty}^{\infty}dq\;qP(q)$, agrees with a direct calculation of $\langle b+b^{\dagger}\rangle$ and is given by $-(2g^2/\hbar\omega_0)(\langle n\rangle -1)$ \cite{Hewson10}. The effective phonon potential shows an asymmetric double well structure, with one of the minima disappearing with increased doping. Near the lowest doping value ($x = 0.4$), the potential
exhibits pronounced anharmonicity, which eventually decreases as the doping increases. For the highest doping level, the shallow minimum has disappeared, and fluctuations decrease, indicated by the presence of a single minimum in the effective potential. The enhanced fluctuations near the instability point are a signal that an instability is imminent.


\end{document}